\documentclass[pra,twocolumn,showpacs]{revtex4}

\usepackage{epsfig,graphicx,tabularx}
\usepackage{graphicx}
\usepackage{graphics}
\usepackage{epsfig}
\usepackage{amsmath,amssymb,mathrsfs}
\usepackage{amsmath}
\usepackage{amsfonts}
\usepackage{amssymb}
\usepackage{verbatim,color,ulem}

\newcommand{\beq}{\begin{equation}}
\newcommand{\eeq}{\end{equation}}
\newcommand{\beqa}{\begin{eqnarray}}
\newcommand{\eeqa}{\end{eqnarray}}

\begin{document}

\title{Electrodynamics of superconductors: \\
from Lorentz to Galilei at zero temperature}

\author{Luca Salasnich$^{1,2,3}$}

\address{$^{1}$Dipartimento di Fisica e Astronomia 
  "Galileo Galilei" and Padua QTech Center,
  Universita di Padova, Via Marzolo 8, 35131, Padova, Italy 
\\
$^{2}$Istituto Nazionale di Fisica Nucleare, Sezione di Padova, 
Via Marzolo 8, 35131, Padova, Italy 
\\
$^{3}$Istituto Nazionale di Ottica del Consiglio Nazionale delle Ricerche,  
Via Nello Carrara 2, 50127 Sesto Fiorentino, Italy}

\begin{abstract}
{We discuss the derivation of the electrodynamics 
of superconductors coupled to the electromagnetic field 
from a Lorentz-invariant bosonic model of Cooper pairs.
Our results are obtained at zero temperature where, according
to the third law of thermodynamics, the entropy of the system is zero.
In the nonrelativistic limit we obtain a Galilei-invariant 
superconducting system which differs with respect to the familiar 
Schr\"odinger-like one. From this point of view, there are 
similarities with the Pauli equation of fermions 
which is derived from the Dirac equation in the nonrelativistic 
limit and has a spin-magnetic field term in contrast with 
the Schr\"odinger equation. One of the peculiar effects of our model is the 
decay of a static electric field inside a superconductor 
exactly with the London penetration length. In addition, 
our theory predicts a modified D'Alembert equation 
for the massive electromagnetic field also in the case 
of nonrelativistic superconducting matter. 
We emphasize the role of the Nambu-Goldstone phase field 
which is crucial to obtain the collective modes of the 
superconducting matter field. In the special case of a 
nonrelativistic neutral superfluid we find a gapless Bogoliubov-like 
spectrum, while for the charged superfluid we obtain a dispersion 
relation that is gapped by the plasma frequency.}
\end{abstract}

\maketitle

\section{Introduction}

There is a renewed interest in the phenomenological description 
of the superconductive electrodynamics taking explicitly in account 
relativistic effects 
\cite{stenuit2001,hirsch2004,tajmar2008,hirsch2015,grigorishin2021} 
or the crucial role of the Nambu-Goldstone phase 
field \cite{nambu1960,goldstone1961} 
writing low-frequency and long-wavelength Lagrangians 
for neutral and charged fermionic superfluids 
\cite{popov1972,popov-book,witten1989,schakel1990,schakel1994,
zhu1995,son2006,schakel-book,sala2008,sala2009,sala2013,tureci2023}. 
Quite surprisingly, in the relativistic models of Refs. 
\cite{stenuit2001,hirsch2004,tajmar2008,hirsch2015,grigorishin2021} 
the nonrelativistic limit of the relativistic matter field was not considered, 
somehow forgetting that the electrons, and the Cooper pairs, 
move at nonrelativistic velocities in the experimentally 
measured superconducting materials
on earth \cite{mermin-book,annett-book,degennes-book,ketterson-book}. 

In this paper we fill this gap by investigating the nonrelativistic 
limit of a relativistic phenomenological model of bosonic 
Cooper pairs minimally coupled to the electromagnetic field. 
Quite remarkably, from the initial Lorentz-invariant setting 
we obtain a Galilei-invariant theory for the superconducting 
matter field which cointains a crucial electromagnetic coupling 
term that is absent in the standard minimally-coupled 
nonrelativistic Schr\"odinger field. This is exactly 
the analog of the coupling between the spin and the magnetic field 
one finds in the Pauli equation, which can be derived 
from the Dirac equation in the nonrelativistic limit \cite{drell-book}. 
By using our improved nonrelativistic formulation of the charged 
matter field, and explicitly taking into account 
the role of the Nambu-Goldstone phase field, working at zero
temperature, where the entropy of the system is also zero, 
we predict effects which should be measurable at very low temperatures,
close to the absolute zero. Some of them have been previously 
suggested \cite{stenuit2001,hirsch2004,tajmar2008,hirsch2015,
grigorishin2021}, but only assuming a quite-unphysical 
relativistic matter field inside the superconductor. 
In particular, we suggest the decay of a static electric field inside 
a superconductor exactly with the London penetration depth. In addition, 
we obtain a modified D'Alembert equation for the massive electromagnetic 
waves inside the nonrelativistic superconducting matter. We finally 
derive a gapped spectum for the density oscillations of the 
charged superfluid made of Cooper pairs. { It is important
  to stress that many classical well-known experimental and theoretical
  results of superconductivity
  \cite{mermin-book,annett-book,degennes-book,ketterson-book},
such as the spontaneous
symmetry breaking of gauge invariance, the London penetration depth
of the magnetic field, and the collective modes of neutral
superfluids, are fully recovered by our formalism.}

\section{Relativistic Cooper pairs and minimal coupling}

We assume that at zero temperature the Cooper pairs in a superconductor 
are described by a relativistic Klein-Gordon \cite{klein1926,gordon1926} 
complex scalar field $\varphi({\bf r},t)$ with Lagrangian density 
\beq
\mathscr{L}_{\rm 0} = {\hbar^2\over 2mc^2} |\partial_t\varphi|^2 - 
        {\hbar^2\over 2m} |{\boldsymbol\nabla}\varphi|^2
- {mc^2\over 2} |\varphi|^2 
        - {\cal E}(|\varphi|^2) \; , 
\label{lstart}
\eeq
where $m=2m_e$ is the mass of a Cooper pair with $m_e$ the electron 
mass, $\hbar$ is the reduced Planck constant, and $c$ is the speed 
of light in vacuum. Here ${\cal E}(|\varphi|^2)$ is the bulk internal energy 
of the system. The Lagrangian (\ref{lstart}) is invariant 
with respect to Lorentz transformations. A similar model was developed by 
Govaerts, Bertrand, and Stenuit \cite{stenuit2001} and by 
Grigorishin \cite{grigorishin2021}. However, 
in Refs. \cite{stenuit2001} and \cite{grigorishin2021} the system is 
supposed to be close to the critical temperature $T_c$, with the 
temperature-dependent 
internal energy ${\cal E}(|\varphi|^2)$ given by the familiar quadratic-quartic 
Mexican hat potential. Here, instead, we work at zero temperature 
and, contrary to the previous papers, we want to emphasize the 
emerging properties in the nonrelativistic limit, where we obtain Maxwell-Proca 
equations \cite{stenuit2001,grigorishin2021,tajmar2008} 
for the electromagnetic field coupled to the nonrelativistic 
superconducting matter. Quite remarkably, we find that 
the coupling between the electromagnetic field and the 
nonrelativistic matter contains a term that is absent { by applying}  
the minimal coupling to the electromagnetic field directly 
into a Schr\"odinger Lagrangian density. Remind that in the 
nonrelativistic limit from the Dirac equation of fermions one gets 
the Pauli equation (which has the spin) and not the Schr\"odinger 
equation (which does not have the spin) \cite{drell-book}. 
What is found here is the bosonic analog of that phenomenon. 

The Cooper pair has the electric charge $q=-2e$ with $e>0$ the modulus 
of the electron charge. The coupling with the electromagnetic field 
is obtained with the minimal substitution 
\beqa 
{\partial_t}&\to& {\partial_t} + i {q\over \hbar} \Phi
\\
{\boldsymbol \nabla} &\to& {\boldsymbol \nabla} - i {q\over \hbar} {\bf A}
\eeqa
where $\Phi({\bf r},t)$ is the electromagnetic scalar potential 
and ${\bf A}({\bf r},t)$ is the electromagnetic vector potential, such that 
\beqa
{\bf E} &=& - {\boldsymbol\nabla}\Phi - \partial_t {\bf A} 
\label{eanda}
\\
{\bf B} &=& {\boldsymbol \nabla} \wedge {\bf A} 
\label{banda}
\eeqa
with ${\bf E}({\bf r},t)$ the electric field and ${\bf B}({\bf r},t)$ 
the magnetic field. 

In this way, the total Lagrangian density 
$\mathscr{L}_{\rm tot}$ of the system is given by 
\beqa 
\mathscr{L}_{\rm tot} = \mathscr{L}_{\rm{shift}} + \mathscr{L}_{\rm{em}} \; , 
\label{kari}
\eeqa
where 
\beqa 
\mathscr{L}_{\rm{shift}} &=& {\hbar^2\over 2mc^2} 
|(\partial_t+ i {q\over \hbar} \Phi) \varphi|^2 - 
{\hbar^2\over 2m} |({\boldsymbol\nabla}- i {q\over \hbar} {\bf A})\varphi|^2 
\nonumber 
\\
&-& {mc^2\over 2} |\varphi|^2 - {\cal E}(|\varphi|^2)
\nonumber 
\\
&=&  \mathscr{L}_{\rm 0} + {q^2\over 2mc^2} |\varphi|^2 \Phi^2 
- {iq\hbar\over 2mc^2} (\varphi^*\partial_t\varphi - 
\varphi\partial_t\varphi^*) \Phi 
\nonumber 
\\
&-& {mc^2\over 2} |\varphi|^2 
- {q^2\over 2m} |\varphi|^2 {\bf A}^2 + {iq\hbar\over 2m} 
(\varphi^* {\boldsymbol\nabla}\varphi 
- \varphi{\boldsymbol\nabla}\varphi^*)\cdot {\bf A} 
\label{kari1}
\eeqa
is the shifted Lagrangian density of relativistic Cooper pairs and 
\beq 
\mathscr{L}_{\rm{em}} = {\epsilon_0\over 2} {\bf E}^2 
- {1\over 2\mu_0} {\bf B}^2  
\label{kari2}
\eeq
is the Lagrangian density of the free electromagnetic field, with 
$\epsilon_0$ the dielectric constant in the vacuum and $\mu_0$ the 
paramagnetic constant in the vacuum. Remember that 
$c=1/\sqrt{\epsilon_0\mu_0}$. The Lagrangian (\ref{kari})
is invariant with respect the local $U(1)$ gauge transformation
$\varphi({\bf r},t) \to \varphi({\bf r},t) \ e^{i\alpha({\bf r},t)}$. 

As well known, the total Lagrangian density (\ref{kari}) can be used to develop 
a zero-temperature quantum field theory by introducing 
the real-time partition function 
\beq 
{\cal Z} = \int {\cal D}[\varphi,\Phi,{\bf A}] \ e^{{i\over \hbar} 
\int dt \, d^3{\bf r} \, \mathscr{L}_{\rm{tot}}} 
\eeq
of the system, within a functional integral 
formalism \cite{popov-book,schakel-book}. The theory can also
be extended at finite temperature by performing a Wick rotation
from real to imaginary time. However, in this paper we do not consider
the finite-temperature effects of the entropy-dependent normal component
of the charged superfluid.
{ The quantum expectation value of the scalar field
$\varphi({\bf r},t)$ is defined as
\beq
\langle \varphi \rangle = {1\over {\cal Z}}
\int {\cal D}[\varphi,\Phi,{\bf A}] \ \varphi \ e^{{i\over \hbar} 
\int dt \, d^3{\bf r} \, \mathscr{L}_{\rm{tot}}} \; . 
\eeq
The transition to the superconducting state is the breaking
of the $U(1)$ local gauge invariance, namely $\langle {\varphi}({\bf r},t)
\rangle \neq 0$ \cite{popov-book,schakel-book}. In this paper 
we work within the saddle-point approximation, where
\beq
{\cal Z} \simeq  e^{{i\over \hbar} 
\int dt \, d^3{\bf r} \, \mathscr{L}_{\rm{tot}}} 
\eeq
and the fields are the ones which extermize the action functional
\beq
S_{\rm{tot}} = \int dt \, d^3{\bf r} \, \mathscr{L}_{\rm{tot}} \; . 
\eeq
In this mean-field framework $\langle {\varphi}({\bf r},t)
\rangle = {\varphi}({\bf r},t)$ and quantum fluctuations are not taken
into account.}

\section{From Lorentz to Galilei} 

The standard way to obtain a Galilei-invariant Schr\"odinger 
matter field $\psi({\bf r},t)$ from the Lorentz-invariant 
Klein-Gordon field $\varphi({\bf r},t)$ is to set 
\beq 
\varphi({\bf r},t) = \psi({\bf r},t) \, e^{-imc^2t/\hbar} \; .
\eeq
Inserting this ansatz into (\ref{lstart}) and (\ref{kari1}) we find 
\beqa 
\mathscr{L}_{\rm 0} &=& {\hbar^2\over 2mc^2}|\partial_t\psi|^2 +
        {i\hbar\over 2} 
(\psi^*\partial_t\psi - \psi\partial_t\psi^*) 
\nonumber 
\\
&-& {\hbar^2\over 2m} |{\boldsymbol\nabla}\psi|^2 - {\cal E}(|\psi|^2) 
\eeqa
and 
\beqa
\mathscr{L}_{\rm{shift}} &=& \mathscr{L}_{\rm 0} + 
{q^2\over 2mc^2} |\psi|^2 \Phi^2 - q |\psi|^2 \Phi 
\nonumber 
\\
&-& {iq\hbar\over 2mc^2} (\psi^*\partial_t\psi - 
\psi\partial_t\psi^*) \Phi - {q^2\over 2m} |\psi|^2 {\bf A}^2 
\nonumber 
\\
&+& {iq\hbar\over 2m} 
(\psi^* {\boldsymbol\nabla}\psi - 
\psi{\boldsymbol\nabla}\psi^*)\cdot {\bf A} 
\eeqa

At this point the matter Lagrangian density is still Lorentz invariant. 
However, under the assumption 
\beq
{\hbar^2\over 2mc^2}|\partial_t\psi|^2 
\ll {i\hbar\over 2} 
(\psi^*\partial_t\psi - \psi\partial_t\psi^*) 
\eeq
we obtain the approximated nonrelativistic Galilei-invariant Lagrangians 
\beq 
\tilde{\mathscr{L}}_{\rm 0} = {i\hbar\over 2} 
(\psi^*\partial_t\psi - \psi\partial_t\psi^*) - 
{\hbar^2\over 2m} |{\boldsymbol\nabla}\psi|^2 - {\cal E}(|\psi|^2) 
\label{don1}
\eeq
and 
\beqa
\tilde{\mathscr{L}}_{\rm{shift}} &=& \tilde{\mathscr{L}}_{\rm 0} + 
{q^2\over 2mc^2} |\psi|^2 \Phi^2 - q |\psi|^2 \Phi 
- {q^2\over 2m} |\psi|^2 {\bf A}^2 
\nonumber 
\\
&+& {iq\hbar\over 2m} (\psi^* {\boldsymbol\nabla}\psi - 
\psi{\boldsymbol\nabla}\psi^*)\cdot {\bf A} \; . 
\label{don2}
\eeqa
Notice that in Eq. (\ref{don1}) we have omitted the term $mc^2|\psi|^2/2$ 
which does not affect the dynamics. As expected, (\ref{don1}) is 
the Lagrangian density of a complex Schr\"odinger field $\psi({\bf r},t)$. 
Instead, quite remarkably, Eq. (\ref{don2}) contains the crucial term 
$q^2\Phi^2|\psi|^2/(2mc^2)=\epsilon_0\mu_0q^2\Phi^2|\psi|^2/(2m)$. 
This term is absent { by applying}  
the minimal coupling to the electromagnetic field directly 
into a Schr\"odinger Lagrangian density. 
To better emphasize this relevant result, let us write 
the Euler-Lagrange equation of (\ref{don2}) with respect 
to $\psi^*({\bf r},t)$, which is given by the following 
nonlinear Schr\"odinger equation 
\beqa 
i\hbar (\partial_t + i{q\over \hbar}\Phi) 
\psi &=& \left[ -{\hbar^2\over 2m}(\nabla - i{q\over \hbar}{\bf A})^2 
+ {\mu(|\psi^2|)} \right] \psi 
\nonumber 
\\
&-& {q^2\over m c^2} \Phi^2 \, \psi \; , 
\label{nlse} 
\eeqa
where 
\beq 
\mu(|\psi|^2) = {\partial {\cal E} \over \partial |\psi|^2}(|\psi|^2) \;  
\label{echem}
\eeq
is the chemical potential of the bulk system as a function 
of the local number density $|\psi|^2$. In Eq. (\ref{nlse}) 
it is the last term which makes the nonrelativistic limit 
of the Klein-Gordon equation coupled to the electromagnetic field 
not equivalent to the fully nonrelativistic Schr\"odinger equation 
coupled to the electromagnetic field. 
This phenomenon is the analog of the Dirac equation 
coupled to the electromagnetic field: 
in the non-relativistic limit from the Dirac equation one gets 
the Pauli equation (which has the spin) and not the Schr\"odinger 
equation (which does not have the spin) \cite{drell-book}. Note that the term 
$q^2\Phi^2/(m c^2)= \epsilon_0\mu_0 \Phi^2/m$ can be discarded 
only if $|q\Phi| \ll m c^2$ but in nonrelativistic superconductors 
this is not the case.
%Keeping this term, the Lagrangian
%$\tilde{\mathscr{L}}_{\rm 0}+ \tilde{\mathscr{L}}_{\rm{shift}}$
%of Eqs. (\ref{don1}) and (\ref{don2}) is still invariant
%with respect the local $U(1)$ gauge transformation 
%$\psi({\bf r},t) \to \psi({\bf r},t) \ e^{i\tilde{\alpha}({\bf r},t)}$.

\section{Density-phase Lagrangian}

The Schr\"odinger field $\psi({\bf r},t)$ of Eqs. (\ref{don1}) 
and (\ref{don2}) is the order parameter of nonrelativistic 
Cooper pairs. We now set 
\beq 
\psi = \sqrt{n_s} \, e^{i\theta} \; ,  
\label{complex}
\eeq
where $n_s({\bf r},t)$ is the number density of Cooper pairs of mass 
$m$ and electric charge $q$, while $\theta({\bf r},t)$ is the Nambu-Goldstone 
phase field \cite{nambu1960,goldstone1961}. 

Several authors  \cite{popov1972,popov-book,witten1989,schakel1990,
schakel1994,zhu1995,son2006,schakel-book,tureci2023} adopted 
the idea of writing a low-frequency and long-wavelength 
Lagrangian density $\mathscr{L}$ of a 
nonrelativistic superfluid in terms of $\theta({\bf r},t)$. 
In all these approaches 
the local superfluid velocity field ${\bf v}_s({\bf r}, t)$
is related to $\theta({\bf r}, t)$ by the fundamental relationship 
\beq 
{\bf v}_s = {\hbar\over m} {\boldsymbol\nabla}\theta \; .  
\label{vs}
\eeq
This equation ensures that the fluid is irrotational, i.e. 
${\boldsymbol\nabla}\wedge {\bf v}_s={\bf 0}$ 
apart from a set of zero measure of quantized vortices. 
Indeed in the presence of a quantized vortex with integer quantum number 
$\kappa$, around it the circulation of the superfluid velocity is such 
that \cite{landau-book,leggett-book} 
\beq 
\oint {\bf v}_s \cdot d{\bf r} = 
{\hbar \over m} \oint {\boldsymbol\nabla}\theta \cdot d{\bf r} 
= {\hbar \over m} \int_0^{{2\pi} \kappa} d\theta = 
{2\pi \hbar \over m} \kappa \; .  
\eeq

Inserting Eq. (\ref{complex}) into 
Eqs. (\ref{don1}) and (\ref{don2}) we obtain the following 
total Lagrangian density
\beq 
\mathscr{L}_{\rm tot} = 
\tilde{\mathscr{L}}_{\rm 0} + \mathscr{L}_{\rm em} + 
\tilde{\mathscr{L}}_{\rm I} \; , 
\label{ltot}
\eeq
where 
\beq 
\tilde{\mathscr{L}}_{\rm 0} = - n_s \, \hbar {\partial_t \theta}  
- n_s \, {\hbar^2\over 2m} ({\boldsymbol \nabla}\theta)^2  
- {\hbar^2\over 8m} {({\boldsymbol\nabla} n_s)^2\over n_s} 
- {\cal E}(n_s) 
\label{l0}
\eeq
is our nonrelativistic { density-phase} Lagrangian,  
$\mathscr{L}_{\rm em}$ is the Lagrangian of the free electromagnetic field, 
given by Eq. (\ref{kari2}), and 
\beq
\tilde{\mathscr{L}}_{\rm I} = n_s \, \epsilon_0\mu_0 {q^2\Phi^2\over 2m}  
- n_s \, q \Phi - n_s \, {q^2{\bf A}^2\over 2m}  + 
n_s \, q \, {\hbar\over m} {\boldsymbol \nabla}\theta \cdot {\bf A} \;  
\label{lI}
\eeq
is the Lagrangian of the interaction between the Cooper pairs 
and the electromagnetic field. 

We observe that in Eq. (\ref{l0}) it appears the 
von Weizs\"acker-like term \cite{von1935} 
\beq 
\tilde{\mathscr{L}}_{{\rm 0,W}} = 
- {\hbar^2\over 8m} {({\boldsymbol\nabla} n_s)^2\over n_s}
\label{l0w}
\eeq
which takes into account the energy cost due to variations 
of the superfluid density. As we will see, this term 
modifies the dispersion relation of collective modes 
of both neutral and charged superfluids. In a very recent 
paper \cite{tureci2023} it has been suggested that Eq. (\ref{l0w}) 
is crucial to obtain a negative electrohydrostatic pressure between 
superconducting bodies at zero temperature.

\subsection{Including the ion background}

In the ground state of superconductor 
there is a compensation between the negative 
electric charge density $\rho_s=qn_s=-2en_s$ of Cooper 
pairs and the positive electric charge density 
$\rho_{\rm bg}=-q {\bar n}_{\rm bg}=2 e {\bar n}_{\rm bg}$ of the background 
of ions with average number density ${\bar n}_{\rm ions}$. 
Taking into account this fact, similarly to the Jellium model 
of a metallic conductor \cite{mermin-book}, 
the full Lagrangian of our model is given by 
\beq 
\mathscr{L}_{\rm full} = \mathscr{L}_{\rm tot} 
+ \mathscr{L}_{\rm bg} \; , 
\label{lfull}
\eeq
where 
\beq 
\mathscr{L}_{\rm bg} = {\bar n}_{\rm bg} \, q \Phi \; . 
\eeq
Notice that we are assuming that the number density ${\bar n}_{\rm bg}$  
of the ion background is space-time independent. 

For the sake of clarity, we stress that in this nonrelativistic 
framework it is the 
Lagrangian density (\ref{lfull}) that must be used to obtain 
the path-integral partition function 
\beq 
{\cal Z}_{\rm full} 
= \int {\cal D}[n_s,\theta,\Phi,{\bf A}] \ e^{{i\over \hbar} 
\int dt \, d^3{\bf r} \, \mathscr{L}_{\rm{full}}} \; , 
\label{supercomplex}
\eeq
where the functional integration is done with respect 
to the local number density $n_s({\bf r},t)$ of Cooper pairs, the 
Nambu-Goldstone field $\theta({\bf r},t)$, and the electromagnetic 
potentials $\Phi({\bf r},t)$ and ${\bf A}({\bf r},t)$. 

\subsection{Charge density and current density}

It is impossible to calculate analytically Eq. (\ref{supercomplex}). 
However, the saddle-point (mean-field) solution is the set of 
Euler-Lagrange equations which are obtained by extremizing 
the action functional 
{ 
\beq 
S_{\rm full} = \int dt \, d^3{\bf r} \, \mathscr{L}_{\rm{full}} \; . 
\eeq
}
The Euler-Lagrange equations of the { full} Lagrangian (\ref{lfull})
with respect to the scalar potential $\Phi({\bf r},t)$ and the vector 
potential ${\bf A}({\bf r},t)$ are nothing else than the Maxwell equations 
\beqa
{\boldsymbol\nabla} \cdot {\bf E} &=& {\rho\over \epsilon_0} 
\label{max1}
\\
{\boldsymbol\nabla} \cdot {\bf B} &=& 0 
\label{max2}
\\
{\boldsymbol\nabla} \wedge {\bf E} &=& - {\partial_t {\bf B}}
\label{max3}
\\
{\boldsymbol\nabla} \wedge {\bf B} &=& \mu_0 \ {\bf j} + 
{\epsilon_0 \mu_0} \ \partial_t{\bf E} 
\label{max4}
\eeqa
where the expressions of the local charge density $\rho({\bf r},t)$ 
and the local current density ${\bf j}({\bf r},t)$ are given by 
\beqa
{\rho} &=& - {\partial \left( \tilde{\mathscr{L}}_{\rm I} 
{ + \mathscr{L}_{\rm bg}} \right) \over \partial \Phi} 
\label{rho}
\\
{\bf j} &=& {\partial \tilde{\mathscr{L}}_{\rm I}\over \partial {\bf A}} 
\label{j}
\eeqa
Thus, one gets 
\beqa
{\rho} &=& q \, n_s { - q \, {\bar n}_{\rm bg}} 
- \epsilon_0 \, {q^2 n_s \mu_0\over m} \, \Phi  
\label{gulp1}
\\
{\bf j} &=&  q \, n_s {\bf v}_s - {1\over \mu_0} {q^2 n_s \mu_0 \over m} 
\, {\bf A}  \; , \label{gulp2}
\eeqa
where the first term 
\beq 
\rho_s=q \, n_s 
\label{rhos}
\eeq
in Eq. (\ref{gulp1}) is the electric charge density 
of Cooper pairs, { the second term is the electric charge density 
of the ion background, and the third term}  
\beq 
\rho_I=-{\epsilon_0\mu_0 q^2 n_s\over m} \, \Phi
\label{rhoI}
\eeq
in Eq. (\ref{gulp1}) is the interaction charge density due to the coupling 
between the Cooper pairs and the electromagnetic scalar potential $\Phi$. 
Instead, the first term 
\beq 
{\bf j}_s=q \, n_s{\bf v}_s 
\label{js}
\eeq
in Eq. (\ref{gulp2}) is the electric current density 
of Cooper pairs, which contains the superfluid 
velocity ${\bf v}_s$ defined in Eq. (\ref{vs}). The second 
term 
\beq 
{\bf j}_I=-{q^2 n_s\over m} \, {\bf A}
\label{jI}
\eeq
in Eq. (\ref{gulp2}) is nothing 
else than the London current \cite{london1935} due to the interaction 
between Cooper pairs and the electromagnetic vector potential ${\bf A}$. 
In 1935 Fritz and Heinz London \cite{london1935} introduced a nonrelativistic 
model where the interaction density $\rho_I$ was included in the total 
electric charge density $\rho$. However, due to the lack of experimental 
evidences \cite{london1936}, subsequently Fritz London discarded  
this term in his book \cite{london-book}. In recent years, 
it has been suggested by Hirsch within an alternative model 
\cite{hirsch2004,hirsch2015} that, at very low temperatures, 
the interaction density $\rho_I$ could be effective and measurable. 

We underline that, as well know, manipulating Eqs. 
(\ref{max1}) and (\ref{max4}) 
one finds the continuity equation for the electric charge density $\rho$ 
and the electric current density ${\bf j}$, namely \cite{jackson-book}
\beq 
\partial_t \rho + {\boldsymbol\nabla} \cdot {\bf j} = 0 \; . 
\label{continuity-full}
\eeq
This result will be used later in combination with a similar, but not equal, 
continuity equation for the superconductive charge density $\rho_s$ 
of Cooper pairs and the electric current density ${\bf j}$. 

Eqs. (\ref{max1}) and (\ref{max2}) equipped with 
Eqs. (\ref{gulp1}) and (\ref{gulp2}) are nothing else than the 
Maxwell-Proca equations for the electrodynamics of 
supercoductors previously discussed in Refs. \cite{stenuit2001,tajmar2008}. 
However, in Ref. \cite{stenuit2001} the Maxwell-Proca equations 
are obtained from a finite-temperature relativistic model 
while in Ref. \cite{tajmar2008} these equations 
are heuristically introduced without a derivation. Here we will 
analyzed the consequences of the Maxwell-Proca equations for superconductors 
at zero temperature, where the normal density is absent. Moreover, 
we will investigate the collective modes of charged superfluid.  

Deep inside a superconductor both magnetic 
field ${\bf B}$ and electric field ${\bf E}$ are 
zero \cite{annett-book}. As a consequence, 
from our Eqs. (\ref{max1}) and (\ref{gulp1}) 
it follows that for the ground state, characterized 
by a uniform and constant number density ${\bar n}_s$ of Cooper pairs 
and a { vanishing} electromagnetic potential $\Phi=0$, 
the total electric charge density $\rho$ is zero, namely 
\beq 
0 = q \left({\bar n}_s - {\bar n}_{\rm bg} \right)  \; ,   
\label{dilanonce}
\eeq 
and consequently ${\bar n}_s={\bar n}_{\rm bg}$. As previously 
discussed, the ion background neutralizes the system.

\subsection{London penetration depth for the static magnetic field}

As discussed above, Eq. (\ref{gulp2}) was obtained 
for the first time by the London brothers \cite{london1935} and it 
gives rise to the expulsion of a magnetic field from a superconductor 
(Meissner-Ochsenfeld effect) \cite{meissner1933}. 

In a static configuration with a zero superfluid velocity ${\bf v}_s$ 
and in the absence of the electric field, i.e. 
${\bf E}={\bf 0}$, the curl of Eq. (\ref{max4}) gives 
\beq 
- \nabla^2 {\bf B} = \mu_0 \ {\boldsymbol \nabla} 
\wedge \left( - {q^2 n_s\over m} {\bf A} \right)  \; , 
\label{crucial}
\eeq
taking into account that 
\beq 
{\boldsymbol\nabla} \wedge ({\boldsymbol \nabla} \wedge {\bf B})= 
- \nabla^2 {\bf B} + {\boldsymbol \nabla} ({\boldsymbol \nabla} \cdot 
{\bf B}) = - \nabla^2 {\bf B} 
\eeq
due to the Gauss law, Eq. (\ref{max2}). Assuming that 
the local density $n_s({\bf r})$ is uniform, i.e. $n_s({\bf r})={\bar n}_s$, 
by using Eq. (\ref{banda}) we get 
\beq 
\nabla^2 {\bf B} = {q^2 {\bar n}_s \mu_0 \over m} {\bf B} \; .
\eeq
Choosing the magnetic field as ${\bf B} = B(x)\, {\bf u}$, with ${\bf u}$ 
a unit vector, the previous equation can be written as 
\beq 
{\partial^2\over \partial x^2} B = 
{q^2 \bar{n}_s \mu_0 \over m} B 
\eeq
which has the following physically relevant solution for 
a superconducting slab defined in the region $x\geq 0$:  
\beq 
B(x) = B(0) \ e^{-x/\lambda_L} \; , 
\label{mah}
\eeq
where 
\beq 
\lambda_L = \sqrt{m\over q^2 \bar{n}_s \mu_0} 
\label{lambdaL}
\eeq
is the so-called London penetration depth, which is typically 
around $100$ nanometers \cite{annett-book}. 
The meaning of Eq. (\ref{mah}) is that inside a superconductor 
the static magnetic field decays exponentially. This is the 
Meissner-Ochsenfeld effect: the expulsion of a magnetic field from a 
superconductor, experimentally observed for the first time 
in 1933 \cite{meissner1933}. 

\subsection{London penetration depth for the static electric field}

It is well know that normal metals screen an external electric 
field ${\bf E}$, which can penetrate at most few 
angstr\"oms (Thomas-Fermi screening length) \cite{mermin-book}. 
For superconducting materials, our equations (\ref{max1}), (\ref{max2}), 
(\ref{gulp1}), and (\ref{gulp2}) suggest that the electric 
field ${\bf E}$ exponentially decays inside a zero-temperature 
superconductor with the much larger London penetration depth $\lambda_L$. 
Let us { show} how to derive this relevant result within our theoretical 
framework. 

In a static configuration, in the absence the  
magnetic field, i.e. {\bf B}={\bf 0}, and assuming a uniform 
number density, the gradient of Eq. (\ref{max1}), with Eq. (\ref{gulp1}) 
and Eq. (\ref{lambdaL}), gives 
\beq 
\nabla^2 {\bf E} = - {1\over \lambda_L^2} {\boldsymbol\nabla} \Phi  \; , 
\label{new-dueto}
\eeq
taking into account that 
\beq 
{\boldsymbol\nabla} ({\boldsymbol \nabla} \cdot {\bf E})= 
\nabla^2 {\bf E} - {\boldsymbol \nabla} \wedge 
({\boldsymbol \nabla} \wedge {\bf E}) = \nabla^2 {\bf E} \; . 
\eeq
{ Notice that to get Eq. (\ref{new-dueto}) it is crucial to 
assume an uniform background $n_{\rm bg}$}. 
In addition, due to Eq. (\ref{eanda}) we find 
\beq 
\nabla^2 {\bf E} = {1\over \lambda_L^2} {\bf E}  \; .  
\eeq
Choosing ${\bf E} = E(x)\, {\bf u}$, with ${\bf u}$ a unit vector, 
the previous equation can be written as 
\beq 
{\partial^2\over \partial x^2} E = {1\over \lambda_L^2} E 
\eeq
which has the following physically relevant solution for 
a superconducting slab defined in the region $x\geq 0$: 
\beq 
E(x) = E(0) \ e^{-x/\lambda_L} \; . 
\label{rimah}
\eeq
The meaning of Eq. (\ref{rimah}) is that inside a zero-temperature 
superconductor the static electric field decays exponentially 
with a characteristic decay length that is exactly the London 
penetration depth $\lambda_L$. 

\subsection{Modified D'Alembert equation for electromagnetic waves}

We investigate what happens to an electromagnetic wave when it is 
suddenly applied to a superconductor in its ground state. 
In full generality, from the Maxwell equations 
(\ref{max1}), (\ref{max2}), (\ref{max3}), and (\ref{max4}) one obtains 
the inhomogeneous wave equations \cite{jackson-book}
\beqa 
\left( {1\over c^2} {\partial^2\over \partial t^2} - \nabla^2 \right) {\bf E} 
&=& - {1\over \epsilon_0} {\boldsymbol\nabla} \rho - \mu_0 \partial_t {\bf j}
\label{tele1bef}
\\
\left( {1\over c^2} {\partial^2\over \partial t^2} - \nabla^2 \right) 
{\bf B} &=& \mu_0 {\boldsymbol\nabla} \wedge {\bf j} \; . 
\label{tele2bef}
\eeqa
Under the assumption that the local number density of Cooper pairs remains 
approximately constant and uniform, i.e. $n_s({\bf r},t)\simeq {\bar n}_s$, 
and with a zero superfluid velocity, 
i.e. ${\bf v}_s({\bf r},t) \simeq {\bf 0}$, 
after remembering Eqs. (\ref{eanda}), (\ref{banda}),  (\ref{rho}), 
and (\ref{j}), from Eqs. (\ref{tele1bef}) and (\ref{tele2bef}) we obtain
\beqa 
\left( {1\over c^2} {\partial^2\over \partial t^2} - \nabla^2 + 
{1\over \lambda_L^2} \right) {\bf E} = {\bf 0}
\label{tele1}
\\
\left( {1\over c^2} {\partial^2\over \partial t^2} - \nabla^2 + 
{1\over \lambda_L^2} \right) 
{\bf B} = {\bf 0}
\label{tele2}
\eeqa
that are the modified D'Alembert equation for the electromagnetic waves 
inside the superconductor with $\lambda_L$ is the London penetration 
depth of Eq. (\ref{lambdaL}). 

The Fourier transform of Eqs. (\ref{tele1}) and (\ref{tele2}) in the 
frequency-wavevector domain $(\omega,{\bf k})$ gives the dispersion relation 
\beq 
\omega = \sqrt{\omega_p^2 + c^2 k^2} \; , 
\label{veromistico}
\eeq
where 
\beq 
\omega_p = {c\over \lambda_L} = \sqrt{q^2{\bar n}_s\over m \epsilon_0} 
\label{omegap}
\eeq
is the Plasma frequency \cite{mermin-book}. 
Thus, the photon spectrum becomes gapped or, in other words, the photon 
acquires a mass. This is nothing else than the Anderson-Higgs 
mechanism \cite{anderson1962,higgs1964,englert1964}, which survives 
in our model also in the context of nonrelativistic superconducting matter. 
Notice that Eq. (\ref{veromistico}) appears also in Refs. 
\cite{grigorishin2021,hirsch2004}. 
As discussed in Ref. \cite{grigorishin2021}, the dispersion relation 
(\ref{veromistico}) can be also written as 
\beq 
k = {1\over c}\sqrt{\omega^2 - \omega_p^2} = \sqrt{{\omega^2\over c^2} 
-{1\over \lambda_L^2}} \; . 
\eeq
Consequently, the electromagnetic plane wave, that is 
proportional to $e^{i({\bf k}\cdot {\bf r}-\omega t)}$ propagates without 
dissipation inside the superconductor for $\omega>\omega_p$. Instead, 
for $\omega <\omega_p$ the electromagnetic wave is damped 
as $e^{- {\bf u} \cdot {\bf r}/\lambda_{\omega}} \, e^{-i\omega t}$ in the interior of 
the superconductor, where ${\bf k}=i {\bf u}/\lambda_{\omega}$ with ${\bf u}$ 
a unit vector and 
\beq 
\lambda_{\omega} = {\lambda_L \over 
\sqrt{1-\left({\omega\over\omega_p}\right)^2}} 
\eeq
is the frequency-dependent penetration depth. Clearly, 
$\lambda_{\omega}\to \lambda_L$ as $\omega\to 0$. Moreover, 
indicating with $\Delta(0)$ the energy gap of Cooper pairs at zero 
temperature, for $\omega > 2\Delta(0)/\hbar$ the charged superfluid 
becomes a normal charged fluid due to the breaking of Cooper pairs. 

\section{Euler-Lagrange equations of superconductors} 

The Euler-Lagrange equation of the { full} Lagrangian 
{ (\ref{lfull})}  
with respect to the Nambu-Goldstone field $\theta({\bf r},t)$ reads
\beq 
\partial_t n_s + {\boldsymbol\nabla} \cdot 
\left( n_s {\bf v}_s - {q n_s\over m} {\bf A} \right) = 0 \; . 
\label{el1}
\eeq
This is nothing else than the continuity equation 
\beq 
\partial_t \rho_s + {\boldsymbol\nabla} \cdot {\bf j} = 0 \; ,  
\label{continuity-super}
\eeq
where the local superconducting charged density $\rho_s({\bf r},t)$ 
is given by Eq. (\ref{rhos}) and the local charged current 
density ${\bf j}({\bf r},t)$ is given by Eq. (\ref{gulp2}). 
Comparing Eq. (\ref{continuity-super}) with Eq. (\ref{continuity-full}) 
it follows that 
\beq 
\partial_t ({\rho}-\rho_s) = 0  \; , 
\label{bufala}
\eeq
i.e. the interaction charge density $\rho_I=\rho-\rho_s$ given 
by Eq. (\ref{rhoI}) must be time independent or, equivalently 
\beq 
\Phi \, \partial_t n_s = - n_s \, \partial_t\Phi \; . 
\eeq

Instead, the Euler-Lagrange equation for the local number density 
$n_s({\bf r},t)$ leads to
\beqa 
\hbar \, \partial_t \theta &+& q \Phi - {\epsilon_0\mu_0q^2\over 2m} \Phi^2 
- {q\over m} {\boldsymbol\nabla}\theta \cdot {\bf A} + {q^2{\bf A}^2\over 2m} 
+ {\hbar^2\over 2m}({\boldsymbol\nabla}\theta)^2 
\nonumber 
\\
&+& {\partial {\cal E} \over \partial n_s} 
-  {\hbar^2\over 2m \sqrt{n_s}} \nabla^2 \sqrt{n_s} = 0 \; .   
\label{careful}
\eeqa
{ By applying} the gradient operator $\boldsymbol\nabla$ 
to Eq. (\ref{careful}) one finds
\beqa 
m \partial_t {\bf v}_s &+& 
{\boldsymbol\nabla} \big[ {1\over 2} m {\bf v}_s^2 + \mu(n_s) 
- {\hbar^2\over 2m \sqrt{n_s}} \nabla^2 \sqrt{n_s} 
+ q \Phi 
\nonumber
\\
&-&{\epsilon_0\mu_0 q^2\over 2m} \Phi^2 
- {q\over \hbar} {\bf v}_s \cdot {\bf A} 
+ {q^2{\bf A}^2\over 2m}\big] = {\bf 0} \; ,  
\label{el2}
\eeqa
where $\mu(n_s)$, given by Eq. (\ref{echem}), is the chemical potential 
of the bulk system as a function of the local number density $n_s({\bf r},t)$. 

\subsection{Gapless collective modes of neutral superfluids} 

In the very special case of a neutral superfluid, i.e. if $q=0$, 
the previous equations become much simpler and it is quite 
easy to determine the collective modes of the zero-temperature 
neutral superfluid. We set 
\beqa 
n_s({\bf r},t) = {\bar n}_s + {\delta n}_s({\bf r},t)
\label{lin1}
\\
{\bf v}_s({\bf r},t) = {\bf 0} + {\delta {\bf v}}_s({\bf r},t)
\label{lin2}
\eeqa
assuming that ${\delta n}_s({\bf r},t)$ and $ {\delta {\bf v}}_s({\bf r},t)$
are small perturbations with respect to the ground-state configuration 
with uniform number density ${\bar n}_s$ and zero superfluid velocity. 

Under the condition $q=0$, the linearized version of 
Eqs. (\ref{el1}) and (\ref{el2}) are then given by 
\beqa 
\partial_t {\delta n}_s + {\bar n}_s {\boldsymbol\nabla} 
\cdot {\delta {\bf v}}_s &=& 0 \; , 
\label{quasi1q0}
\\
{\bar n}_s \partial_t {\delta {\bf v}}_s + c_s^2 {\boldsymbol\nabla} 
{\delta n}_s - {\hbar^2\over 4m^2} {\boldsymbol\nabla} 
(\nabla^2{\delta n}_s) &=& {\bf 0} \; , 
\label{quasi2q0}
\eeqa
where 
\beq 
c_s = \sqrt{ {\bar{n}_s\over m}{\partial \mu\over \partial n}({\bar n}_s)}
\eeq 
is the speed of sound.
{ By applying} the time derivative $\partial_t$ to Eq. (\ref{quasi1q0}) 
and the divergence ${\boldsymbol\nabla}\cdot$ to Eq. (\ref{quasi2q0}) 
and subtracting the two resulting equations we find 
\beq 
\left( \partial_t^2  - c_s^2 \nabla^2 + 
{\hbar^2\over 4m^2} \nabla^4 \right) {\delta n}_s = 0 \; ,  
\label{semprepiuq0}
\eeq
The Fourier transform of Eq. (\ref{semprepiuq0}) in the 
frequency-wavevector domain $(\omega,{\bf k})$ gives the dispersion relation 
\beq 
\omega = \sqrt{c_s^2 k^2 + {\hbar^2k^4\over 4m^2} } \; . 
\label{mainq0}
\eeq
This dispersion relation is a gapless Bogoliubov-like 
spectrum \cite{bogoliubov1947}, which reduces to the phonon spectrum 
\beq 
\omega=c_sk
\eeq
at very low wavenumbers. while for large wavenumbers one finds 
\beq 
\omega = {\hbar k^2\over 2m}
\eeq
that is the single-particle spectrum of free massive particles. 
For the sake of completeness, we underline that Eq. (\ref{mainq0}) 
is fully consistent with our previous results \cite{sala2008,sala2009,sala2013}
for the collective modes of nonrelativistic neutral fermionic superfluids 
with the inclusion of the von Weizs\"acker-like term, Eq. (\ref{l0w}). 

\subsection{Gapped collective modes of charged superfluids} 

We now analyze the collective modes of a zero-temperature 
superconductor. In this case $q\neq 0$ and, in addition to 
Eqs. (\ref{lin1}) and (\ref{lin2}), we must set 
\beqa 
{\Phi}({\bf r},t) = { 0} + \delta\Phi({\bf r},t)
\\
{\bf A}({\bf r},t) = {\bf 0} + \delta{\bf A}({\bf r},t)
\eeqa
assuming that ${\delta \Phi}({\bf r},t)$ and ${\delta {\bf A}}({\bf r},t)$
are small perturbations with respect to the ground-state 
electromagnetic configuration of { zero} scalar potential 
and zero vector potential.

Under the condition $q\neq 0$, the linearized version 
of Eqs. (\ref{el1}) and (\ref{el2}) are then given by 
\beqa 
\partial_t ({\delta n}_s) &+& {\bar n}_s {\boldsymbol\nabla} 
\cdot {\delta {\bf v}}_s - {q {\bar n}_s\over m} 
{\boldsymbol\nabla}\cdot \delta{\bf A} = 0 \; , 
\label{quasi1}
\\
{\bar n}_s \partial_t ({\delta {\bf v}}_s) &+& c_s^2 {\boldsymbol\nabla}
({\delta n}_s) - {\hbar^2\over 4m^2} {\boldsymbol\nabla} 
(\nabla^2{\delta n}_s) 
\nonumber 
\\
&+& {q{\bar n}_s\over m} {\boldsymbol\nabla}(\delta{\Phi}) = {\bf 0} \; . 
\label{quasi2}
\eeqa
Similarly, the linearized version of the 
Maxwell equations (\ref{max1}), (\ref{max2}), (\ref{max3}), and 
(\ref{max4}) reads 
\beqa
{\boldsymbol\nabla} \cdot \delta{\bf E} &=& {\delta\rho\over \epsilon_0} 
\label{max1-lin}
\\
{\boldsymbol\nabla} \cdot \delta{\bf B} &=& 0 
\label{max2-lin}
\\
{\boldsymbol\nabla} \wedge \delta{\bf E} &=& \partial_t (\delta{\bf B})
\label{max3-lin}
\\
{\boldsymbol\nabla} \wedge \delta {\bf B} &=& \mu_0 \ \delta{\bf j} + 
{\epsilon_0 \mu_0} \ \partial_t(\delta{\bf E}) 
\label{max4-lin}
\eeqa
where 
\beqa
\delta{\bf E} &=& - {\boldsymbol\nabla}(\delta\Phi ) - 
\partial_t(\delta{\bf A})  
\label{eanda-lin}
\\
\delta{\bf B} &=& {\boldsymbol \nabla} \wedge \delta{\bf A}  
\label{banda-lin}
\eeqa
and 
\beqa
\delta{\rho} &=& q \,  
\delta n_s  - \epsilon_0 {q^2 {\bar n}_s \mu_0\over m} \, \delta\Phi 
\label{gulp1-lin}
\\
\delta{\bf j} &=&  q \, {\bar n}_s \delta {\bf v}_s - 
{1\over \mu_0} {q^2\mu_0 \over m} {\bf A} \, \delta n_s 
- {1\over \mu_0} {q^2{\bar n}_s\mu_0 \over m} \, \delta {\bf A} . 
\label{gulp2-lin}
\eeqa

Needless to say, finding analytical solutions of the coupled equations 
from (\ref{quasi1}) to (\ref{gulp2-lin}) seems not easy. However, 
we are able to obtain some interesting result. 
{ By applying} the time derivative $\partial_t$ to Eq. (\ref{quasi1}) 
and the divergence ${\boldsymbol\nabla}\cdot$ to Eq. (\ref{quasi2}) 
and subtracting the two resulting equations we find 
\beq 
\left( \partial_t^2  - c_s^2 \nabla^2 + 
{\hbar^2\over 4m^2} \nabla^4 \right) {\delta n}_s + 
{q{\bar n}_s\over m} {\boldsymbol\nabla}\cdot \delta {\bf E} = 0 \; ,  
\label{semprepiu}
\eeq
taking into account Eq. (\ref{eanda-lin}). 
Then, from the first Maxwell equation (\ref{max1-lin}) and 
Eq. (\ref{gulp1-lin}) we get 
\beq
{\boldsymbol\nabla}\cdot \delta {\bf E} = 
{\delta \rho\over \epsilon_0} = {q\over \epsilon_0} \delta n_s + 
{\delta \rho_I\over \epsilon_0} \; . 
\eeq
Remembering that $\rho_I=\rho-\rho_s$ is constant in time, 
{ as shown} by Eq. (\ref{bufala}), 
{ by applying} the 
operator $\partial_t$ to Eq. (\ref{semprepiu}) we obtain 
\beq 
\left( \partial_t^3  - c_s^2 \partial_t \nabla^2 + 
{\hbar^2\over 4m^2} \partial_t \nabla^4 + \omega_p^2 \partial_t 
\right) {\delta n}_s = 0 \; ,    
\label{semprepiuwrong}
\eeq
which gives the dispersion relation $\omega=0$  but also 
\beq 
\omega = \sqrt{\omega_p^2 + c_s^2 k^2  + {\hbar^2k^4\over 4m^2} } 
\label{main}
\eeq
that is a gapped generalization of Eq. (\ref{mainq0}). As expected,  
the gap is exactly due to the plasma frequency $\omega_p$ of 
Eq. (\ref{omegap}).  

\section{Conclusions}

We have analyzed several consequences 
of a time-dependent relativistic model of bosonic charged 
Cooper pairs minimally coupled to the electromagnetic field. 
While our model shares similarities with other relativistic treatments of 
superconductivity \cite{stenuit2001,hirsch2004,tajmar2008,
hirsch2015,grigorishin2021}, it differs in at least three important ways. 
First, our results have been obtained at zero temperature where, at least 
for clean superconductors, the normal component of the 
superconducting electrons is zero and a real-time description of the 
supercondictive bosonic field is fully justified \cite{varlamov-book}. 
Second, we have explicitly discussed the derivation of the nonrelativistic 
model for the matter field from the relativistic one, emphasizing 
the crucial role of a term which couples the density of Cooper pairs 
with the electromagnetic scalar potential. This term can be also 
obtained \cite{babaev-book} from the nonrelativistic 
low-frequency and long-wavelength Popov's 
action \cite{popov1972,popov-book} of a charged 
superfluid, but only performing a quantum-mechanical functional 
integration with respect to the density field { within the saddle-point 
approximation}. Third, we have obtained 
the full set of equations for nonrelativistic 
charged superfluids coupled to the (relativistic) electromagnetic field 
in terms of the superfluid density and the superfluid velocity, 
that is directly related to the gradient of the 
Nambu-Goldstone phase field. In these equations, in addition 
to the previously discussed coupling term, there is a 
von Weizs\"acker-like term \cite{von1935} which takes into account 
the energy cost due to variations of the superfluid density 
and modifies the dispersion relation of superfluid collective modes. 

Our model supports the claim \cite{hirsch2004,hirsch2015,
grigorishin2021} that, very 
close to zero temperature, it should be possible 
to experimentally measure the decay of a static electric field 
inside a superconductor with a characteristic length that is 
the London penetration depth instead of the Thomas-Fermi screening length. 
A recent experimental attempt to measure this effect 
by using atomic force microscopy on a niobium sample was inconclusive due 
to limited accuracy \cite{peronio2016}. We expect that 
near-future experiments could test also other zero-temperature 
predictions discussed in this paper: a gapped spectrum of the 
electromagnetic waves inside the superconductor and the gapped spectrum 
of the supercondicting density oscillations. To achieve these 
goals it is necessary to work at extremely low temperatures, where 
the normal component, containing the entropy and the viscosity 
of the system, is negligible. This is the main 
experimental problem that needs to be overcome.

\vskip 0.3cm

{The author thanks Andrea Bardin, Luca Dell'Anna, 
Koichiro Furutani, Francesco Lorenzi, Maria Pelizzo, Andrea Perali, and 
Dima Sorokin for useful discussions and suggestions. 
This work is partially supported by the European Union-NextGenerationEU within 
the National Center for HPC, Big Data and Quantum Computing 
[Project No. CN00000013, CN1 Spoke 10: “Quantum Computing”], 
by the BIRD Project "Ultracold atoms in curved geometries" of the 
University of Padova, 
by “Iniziativa Specifica Quantum” of Istituto Nazionale 
di Fisica Nucleare, { and} by the European Quantum
Flagship Project "PASQuanS 2". { The projects  
"Frontiere Quantistiche" (Dipartimenti di Eccellenza) and}
PRIN 2022 "Quantum Atomic Mixtures: Droplets, 
Topological Structures, and Vortices" have been financed by
the Italian Ministry for University and Research and the European
Union-NextGenerationEU.}


\begin{thebibliography}{99}

\bibitem{stenuit2001} J. Govaerts, D. Bertrand, and G. Stenuit, 
Supercond. Sci. Technol. {\bf 14}, 463 (2001). 

\bibitem{hirsch2004} J.E. Hirsch, Phys. Rev. B {\bf 69}, 214515 (2004). 

\bibitem{tajmar2008} M. Tajmar, Phys. Lett. A {\bf 372}, 3289 (2008). 

\bibitem{hirsch2015} J.E. Hirsch, Physica C {\bf 508}, 21 (2015). 

\bibitem{grigorishin2021} K.V. Grigorishin, J. Low Temp. Phys. 
{\bf 203}, 262 (2021). 

\bibitem{nambu1960} Y. Nambu, Phys. Rev. {\bf 117}, 648 (1960). 

\bibitem{goldstone1961} J. Goldstone, Nuovo Cim. {\bf 19}, 154 (1961). 

\bibitem{popov1972} V. N. Popov, Theor. and Math. Phys. 
{\bf 11}, 478 (1972). 

\bibitem{popov-book} V.N. Popov, {\it Functional Integrals in Quantum 
Field Theory and Statistical Physics}, (Reidel, 1983).

\bibitem{witten1989} M. Greiter, F. Wilczek, and E. Witten, 
Mod. Phys. Lett. B {\bf 3}, 903 (1989). 

\bibitem{schakel1990} A.M.J. Schakel, Mod. Phys. Lett. B {\bf 4}, 927 (1990). 

\bibitem{schakel1994} A.M.J. Schakel,  Mod. Phys. Lett. B {\bf 8}, 2021 (1994).

\bibitem{zhu1995} I.J.R. Aitchison, P. Ao, D.J. Thouless, and X.-M. Zhu, 
Phys. Rev. B {\bf 51}, 6531 (1995). 

\bibitem{son2006} D.T. Son and M. Wingate, Ann. Phys. {\bf 321}, 197 (2006). 

\bibitem{schakel-book} A.M.J. Schakel, 
{\it Boulevard of Broken Symmetries} (World Scientific, 2008). 

\bibitem{sala2008} L. Salasnich and F. Toigo, 
Phys. Rev. A {\bf 78}, 053626 (2008). 

\bibitem{sala2009} L. Salasnich, Laser Phys. {\bf 19}, 642 (2009). 

\bibitem{sala2013} L. Salasnich, P. Comaron, M. Zambon, and F. Toigo, 
Phys. Rev. A {\bf 88}, 033610 (2013). 

\bibitem{tureci2023} T.J. Maldonado, D.N. Pham, A. Amaolo, 
A.W. Rodriguez, and H. T\"ureci, e-preprint arXiv:2307.0490. 

\bibitem{mermin-book} N.W. Ashcroft and N.D. Mermin, 
{\it Solid State Physics} (Thompson, 2003). 

\bibitem{annett-book} J. Annett, 
{\it Superconductivity, Superfluids and Condensates} 
(Oxford Univ. Press, 2005).

{ 

\bibitem{degennes-book} P.G. De Gennes,
Superconductivity of Metals and Alloy (Westview Press, 1999). 

\bibitem{ketterson-book} J.B. Ketterson and S.N. Song, 
Superconductivity (Cambridge Univ. Press, 1999). 

}
  
\bibitem{drell-book} J.D. Bjorken and S.D. Drell, 
{\it Relativistic Quantum Mechanics} (McGraw-Hill, 1964). 

\bibitem{klein1926} O. Klein, Z. Phys. {\bf 37}, 895 (1926).

\bibitem{gordon1926} W. Gordon, Z. Phys. {\bf 40}, 117 (1926).

\bibitem{landau-book} L.D. Landau and E.M. Lifshitz, 
{\it Fluid Mechanics}, vol. 6 of Course of Theoretical Physics 
(Pergamon Press, 1987).

\bibitem{leggett-book} A.J. Leggett, {\it Quantum Liquids}  
(Oxford Univ. Press, 2006).

\bibitem{von1935} C.F. von Weizs\"acker, Z. Phys. {\bf 96}, 431 (1935). 

\bibitem{london1935} F. London and H. London, Proc. Roy. Soc. A {\bf 149}, 
71 (1935). 

\bibitem{london1936} H. London, Proc. Roy. Soc. A {\bf 155}, 
102 (1936). 

\bibitem{london-book} F. London, {\it Superfluids} (Dover, 1961). 

\bibitem{jackson-book} J.D. Jackson, {\it Classical Electrodynamics} 
(Wiley, 1998). 

\bibitem{meissner1933} W. Meissner and R. Ochsenfeld, 
Naturwissenschaften {\bf 21}, 787 (1933). 

\bibitem{anderson1962} P.W. Anderson, Phys. Rev. {\bf 130}, 439 (1962). 

\bibitem{higgs1964} P.W. Higgs, Phys. Rev. Lett. {\bf 13}, 508 (1962). 

\bibitem{englert1964} F. Englert and R. Brout, Phys. Rev. Lett. {\bf 13}, 
321 (1964). 

\bibitem{bogoliubov1947} N.N. Bogoliubov, J. Phys. (USSR) {\bf 11}, 23 (1947).

\bibitem{varlamov-book} A. Larkin and A. Varlamov, 
{\it Theory of fluctuations in superconductors} (Clarendon Press, 2007). 

\bibitem{babaev-book} B. Svistunov, E. Babaev, and N. 
Prokof'ev, {\it Superfluid States of Matter} (CRC Press, 2015). 

\bibitem{peronio2016} A. Peronio and F.J. Giessibl, 
Phys. Rev. B {\bf 94}, 094503 (2016). 

\end{thebibliography}
\end{document}